\RequirePackage{lineno}
\documentclass[aps,prl,showpacs,superscriptaddress,twocolumn]{revtex4}  
\usepackage{graphicx}  
\usepackage{dcolumn}   

\usepackage{setspace}  
\usepackage{bm}        
\usepackage{ulem}      
\usepackage{color}     %
\usepackage{amssymb}   
\usepackage{multirow}
\usepackage{epstopdf}  

\newcommand {\et}            {\mbox{$E_T$}}

\newcommand {\gev}           {\mbox{$\rm GeV$}}
\newcommand {\gevc}          {\mbox{${\rm GeV\!/c}$}}
\newcommand {\gevcc}         {\mbox{${\rm GeV}\!/c^2$}}

\newcommand {\gt}            {\mbox{$>$}}

\newcommand {\invfb}            {\ensuremath {{\rm fb}^{-1}}}
\newcommand {\luminosity}    {\mbox{$4.3 \,\invfb$}}
\newcommand {\met}           {\mbox{$\not\!\!\et$}}

\newcommand {\ppbar}         {\mbox{$p\bar{p}$}}

\newcommand {\wenu}          {\mbox{$ W^{\pm} \rightarrow e^{\pm} \nu$}}
\newcommand {\wtaunu}        {\mbox{$ W^{\pm} \rightarrow \tau^{\pm} \nu$}}
\newcommand {\wpigamma}      {\mbox{$ W^{\pm} \rightarrow \pi^{\pm} \gamma$}}
\newcommand {\bratio}        {\mbox{$\Gamma (W^{\pm}\to \pi^{\pm}\gamma)/\Gamma (W^{\pm}\to e^{\pm} \nu)$}} 

\newcommand {\zee}           {\mbox{$ Z^0 \rightarrow e^+ e^-$}}

\newcommand {\Et}            {\ensuremath {E_{T}}}

\begin{document}

\topmargin -0.6in
\textheight 9in

\hspace{5.2in}  

\title{Search for the Rare Radiative Decay: $W\rightarrow\pi\gamma$ in
\ppbar\ Collisions at $\sqrt{s} = 1.96$ TeV}

\affiliation{Institute of Physics, Academia Sinica, Taipei, Taiwan 11529, Republic of China} 
\affiliation{Argonne National Laboratory, Argonne, Illinois 60439, USA} 
\affiliation{University of Athens, 157 71 Athens, Greece} 
\affiliation{Institut de Fisica d'Altes Energies, ICREA, Universitat Autonoma de Barcelona, E-08193, Bellaterra (Barcelona), Spain} 
\affiliation{Baylor University, Waco, Texas 76798, USA} 
\affiliation{Istituto Nazionale di Fisica Nucleare Bologna, $^{aa}$University of Bologna, I-40127 Bologna, Italy} 
\affiliation{University of California, Davis, Davis, California 95616, USA} 
\affiliation{University of California, Los Angeles, Los Angeles, California 90024, USA} 
\affiliation{Instituto de Fisica de Cantabria, CSIC-University of Cantabria, 39005 Santander, Spain} 
\affiliation{Carnegie Mellon University, Pittsburgh, Pennsylvania 15213, USA} 
\affiliation{Enrico Fermi Institute, University of Chicago, Chicago, Illinois 60637, USA}
\affiliation{Comenius University, 842 48 Bratislava, Slovakia; Institute of Experimental Physics, 040 01 Kosice, Slovakia} 
\affiliation{Joint Institute for Nuclear Research, RU-141980 Dubna, Russia} 
\affiliation{Duke University, Durham, North Carolina 27708, USA} 
\affiliation{Fermi National Accelerator Laboratory, Batavia, Illinois 60510, USA} 
\affiliation{University of Florida, Gainesville, Florida 32611, USA} 
\affiliation{Laboratori Nazionali di Frascati, Istituto Nazionale di Fisica Nucleare, I-00044 Frascati, Italy} 
\affiliation{University of Geneva, CH-1211 Geneva 4, Switzerland} 
\affiliation{Glasgow University, Glasgow G12 8QQ, United Kingdom} 
\affiliation{Harvard University, Cambridge, Massachusetts 02138, USA} 
\affiliation{Division of High Energy Physics, Department of Physics, University of Helsinki and Helsinki Institute of Physics, FIN-00014, Helsinki, Finland} 
\affiliation{University of Illinois, Urbana, Illinois 61801, USA} 
\affiliation{The Johns Hopkins University, Baltimore, Maryland 21218, USA} 
\affiliation{Institut f\"{u}r Experimentelle Kernphysik, Karlsruhe Institute of Technology, D-76131 Karlsruhe, Germany} 
\affiliation{Center for High Energy Physics: Kyungpook National University, Daegu 702-701, Korea; Seoul National University, Seoul 151-742, Korea; Sungkyunkwan University, Suwon 440-746, Korea; Korea Institute of Science and Technology Information, Daejeon 305-806, Korea; Chonnam National University, Gwangju 500-757, Korea; Chonbuk National University, Jeonju 561-756, Korea} 
\affiliation{Ernest Orlando Lawrence Berkeley National Laboratory, Berkeley, California 94720, USA} 
\affiliation{University of Liverpool, Liverpool L69 7ZE, United Kingdom} 
\affiliation{University College London, London WC1E 6BT, United Kingdom} 
\affiliation{Centro de Investigaciones Energeticas Medioambientales y Tecnologicas, E-28040 Madrid, Spain} 
\affiliation{Massachusetts Institute of Technology, Cambridge, Massachusetts 02139, USA} 
\affiliation{Institute of Particle Physics: McGill University, Montr\'{e}al, Qu\'{e}bec, Canada H3A~2T8; Simon Fraser University, Burnaby, British Columbia, Canada V5A~1S6; University of Toronto, Toronto, Ontario, Canada M5S~1A7; and TRIUMF, Vancouver, British Columbia, Canada V6T~2A3} 
\affiliation{University of Michigan, Ann Arbor, Michigan 48109, USA} 
\affiliation{Michigan State University, East Lansing, Michigan 48824, USA}
\affiliation{Institution for Theoretical and Experimental Physics, ITEP, Moscow 117259, Russia}
\affiliation{University of New Mexico, Albuquerque, New Mexico 87131, USA} 
\affiliation{Northwestern University, Evanston, Illinois 60208, USA} 
\affiliation{The Ohio State University, Columbus, Ohio 43210, USA} 
\affiliation{Okayama University, Okayama 700-8530, Japan} 
\affiliation{Osaka City University, Osaka 588, Japan} 
\affiliation{University of Oxford, Oxford OX1 3RH, United Kingdom} 
\affiliation{Istituto Nazionale di Fisica Nucleare, Sezione di Padova-Trento, $^{bb}$University of Padova, I-35131 Padova, Italy} 
\affiliation{LPNHE, Universite Pierre et Marie Curie/IN2P3-CNRS, UMR7585, Paris, F-75252 France} 
\affiliation{University of Pennsylvania, Philadelphia, Pennsylvania 19104, USA}
\affiliation{Istituto Nazionale di Fisica Nucleare Pisa, $^{cc}$University of Pisa, $^{dd}$University of Siena and $^{ee}$Scuola Normale Superiore, I-56127 Pisa, Italy} 
\affiliation{University of Pittsburgh, Pittsburgh, Pennsylvania 15260, USA} 
\affiliation{Purdue University, West Lafayette, Indiana 47907, USA} 
\affiliation{University of Rochester, Rochester, New York 14627, USA} 
\affiliation{The Rockefeller University, New York, New York 10065, USA} 
\affiliation{Istituto Nazionale di Fisica Nucleare, Sezione di Roma 1, $^{ff}$Sapienza Universit\`{a} di Roma, I-00185 Roma, Italy} 

\affiliation{Rutgers University, Piscataway, New Jersey 08855, USA} 
\affiliation{Texas A\&M University, College Station, Texas 77843, USA} 
\affiliation{Istituto Nazionale di Fisica Nucleare Trieste/Udine, I-34100 Trieste, $^{gg}$University of Trieste/Udine, I-33100 Udine, Italy} 
\affiliation{University of Tsukuba, Tsukuba, Ibaraki 305, Japan} 
\affiliation{Tufts University, Medford, Massachusetts 02155, USA} 
\affiliation{University of Virginia, Charlottesville, VA  22906, USA}
\affiliation{Waseda University, Tokyo 169, Japan} 
\affiliation{Wayne State University, Detroit, Michigan 48201, USA} 
\affiliation{University of Wisconsin, Madison, Wisconsin 53706, USA} 
\affiliation{Yale University, New Haven, Connecticut 06520, USA} 
\author{T.~Aaltonen}
\affiliation{Division of High Energy Physics, Department of Physics, University of Helsinki and Helsinki Institute of Physics, FIN-00014, Helsinki, Finland}
\author{B.~\'{A}lvarez~Gonz\'{a}lez$^w$}
\affiliation{Instituto de Fisica de Cantabria, CSIC-University of Cantabria, 39005 Santander, Spain}
\author{S.~Amerio}
\affiliation{Istituto Nazionale di Fisica Nucleare, Sezione di Padova-Trento, $^{bb}$University of Padova, I-35131 Padova, Italy} 

\author{D.~Amidei}
\affiliation{University of Michigan, Ann Arbor, Michigan 48109, USA}
\author{A.~Anastassov}
\affiliation{Northwestern University, Evanston, Illinois 60208, USA}
\author{A.~Annovi}
\affiliation{Laboratori Nazionali di Frascati, Istituto Nazionale di Fisica Nucleare, I-00044 Frascati, Italy}
\author{J.~Antos}
\affiliation{Comenius University, 842 48 Bratislava, Slovakia; Institute of Experimental Physics, 040 01 Kosice, Slovakia}
\author{G.~Apollinari}
\affiliation{Fermi National Accelerator Laboratory, Batavia, Illinois 60510, USA}
\author{J.A.~Appel}
\affiliation{Fermi National Accelerator Laboratory, Batavia, Illinois 60510, USA}
\author{A.~Apresyan}
\affiliation{Purdue University, West Lafayette, Indiana 47907, USA}
% ...
\collaboration{CDF Collaboration\footnote{With visitors from $^a$University of Massachusetts Amherst, Amherst, Massachusetts 01003,
$^b$Istituto Nazionale di Fisica Nucleare, Sezione di Cagliari, 09042 Monserrato (Cagliari), Italy,
$^c$University of California Irvine, Irvine, CA  92697, 
$^d$University of California Santa Barbara, Santa Barbara, CA 93106
$^e$University of California Santa Cruz, Santa Cruz, CA  95064,
$^f$CERN,CH-1211 Geneva, Switzerland,
$^g$Cornell University, Ithaca, NY  14853, 
$^h$University of Cyprus, Nicosia CY-1678, Cyprus, 
$^i$University College Dublin, Dublin 4, Ireland,
$^j$University of Fukui, Fukui City, Fukui Prefecture, Japan 910-0017,
$^k$Universidad Iberoamericana, Mexico D.F., Mexico,
$^l$Iowa State University, Ames, IA  50011,
$^m$University of Iowa, Iowa City, IA  52242,
$^n$Kinki University, Higashi-Osaka City, Japan 577-8502,
$^o$Kansas State University, Manhattan, KS 66506,
$^p$University of Manchester, Manchester M13 9PL, England,
$^q$Queen Mary, University of London, London, E1 4NS, England,
$^r$University of Melbourne, Victoria 3010, Australia,
$^s$Muons, Inc., Batavia, IL 60510,
$^t$Nagasaki Institute of Applied Science, Nagasaki, Japan, 
$^u$National Research Nuclear University, Moscow, Russia,
$^v$University of Notre Dame, Notre Dame, IN 46556,
$^w$Universidad de Oviedo, E-33007 Oviedo, Spain, 
$^x$Texas Tech University, Lubbock, TX  79609, 
$^y$Universidad Tecnica Federico Santa Maria, 110v Valparaiso, Chile,
$^z$Yarmouk University, Irbid 211-63, Jordan,
$^{hh}$On leave from J.~Stefan Institute, Ljubljana, Slovenia, 
}}
\noaffiliation
\date{\today, version 1.1}

\begin{abstract}

 We present a search for the rare radiative decay \wpigamma\ using
 data corresponding to an integrated luminosity of 4.3 \invfb\ of
 proton-antiproton collisions at a center of mass energy of 1.96
 TeV collected by the CDF experiment at Fermilab.  As no statistically
 significant signal is observed, we set a 95\% confidence level upper
 limit on the relative branching fraction \bratio\ at $6.4\times
 10^{-5}$, a factor of 10 improvement over the previous limit.

\end{abstract}

\maketitle

Since their discovery by the UA1 \cite{UA1} and UA2 \cite{UA2}
experiments, the properties and decays of the massive mediators of the
electroweak interaction, the $W$ and $Z$ vector bosons, have been
studied extensively. The exclusive leptonic final 
states of the boson decays, $Z\rightarrow l^{+}l^{-}$ and
$W^{\pm}\rightarrow l^{\pm}\nu$, have been observed and are
well-understood. Hadronic decays of the bosons have also been observed, 
but no exclusive final states predicted by the standard model (SM) 
have been identified. 
A class of these unobserved vector boson decay channels, radiative
decays ($V \to P+\gamma$), where $P$ is a pseudoscalar meson, are
sensitive probes of the strong dynamics involved in meson formation as
well as of the vector boson couplings to the photon. These radiative
decays are rare because their branching ratios are suppressed by a
factor of $(f_{P}/M_{W,Z})^2$, where $f_{P}$ is the meson form factor.
Standard model predictions for the branching ratio of the decay of the
$W$ boson to charged pion and photon (\wpigamma\ ) range from $\sim
10^{-6}$ to $\sim 10^{-8}$
\cite{theory}.  However, the sensitivity of the Tevatron experiments
to observing these rare decays, especially
\wpigamma, is approaching the upper range of the SM
predictions. Measurement of the cross section of these radiative
$W$ and $Z$ boson decays would provide information about these poorly
understood SM processes, especially involving meson form factors at high
energies.

Previous searches for the \wpigamma\ decay by the UA1, UA2, and CDF
experiments have successively brought the upper limit on \bratio\ down
to the level of $7 \times 10^{-4}$ at 95\% confidence level
(C.L.)\cite{previous_searches,kadel}.

In this Letter we present a search for the \wpigamma\ decay using data
corresponding to \luminosity\ of integrated luminosity recorded by the
CDF experiment in \ppbar\ collisions at 1.96 TeV.  Compared to the
previous CDF search \cite{kadel}, this represents an increase in the
size of the data set by a factor of $\approx 50$.

The CDF II detector is a general purpose particle detector and is described
in detail elsewhere \cite{CDF}.  The $z$-axis of the detector
coordinate system points along the direction of the proton beam.  The
event geometry and kinematics are described using the azimuthal angle
$\phi$ and the pseudorapidity $\eta=-\ln(\tan\theta/2)$, where
$\theta$ is the polar angle with respect to the beam axis.  The
transverse energy and momentum of the reconstructed particles are
defined as $E_T= E\sin{\theta}$, $p_T =p\sin{\theta}$, where $E$ is
the energy and $p$ is the momentum.

The analysis presented below uses electron ($e^{\pm}$), photon
($\gamma$), and charged pion ($\pi^{\pm}$) candidates, reconstructed
and identified using information from the central outer tracker (COT)
\cite{COT}, central electromagnetic (CEM) and hadronic calorimeters
(CHA), and the central shower maximum detector (CES)
\cite{cdf_cem_ces,cdf_cha}.

Candidate $\pi^{\pm} \gamma$ events are collected using an inclusive
photon trigger that has no tracking requirements. As a result, the
trigger has similar efficiencies for selecting high-$p_T$ electrons
and photons. The kinematic properties of the \wpigamma\ decay are also
in many respects close to the kinematics properties of the \wenu\
decay, which has been extensively studied at the Tevatron
\cite{wenu}. Consequently, many common systematic uncertainties cancel
in the ratio \bratio, and we use sample of \wenu\ events collected with the same trigger to normalize the results of the search. 

The three level calorimeter-based trigger selects events with at least
one central calorimeter cluster with $E_T >$ 25 GeV and a
large fraction of energy deposited in the CEM: $E_{CHA}/E_{CEM} <
0.0055+0.00045\times E$. A cluster is required to be isolated such that the
additional energy in a cone of $\Delta R =
\sqrt{\Delta \phi^2 + \Delta \eta^2} = 0.4$ is less
than 10\% of the cluster energy. To reduce background from the
multi-photon decays of neutral mesons, most importantly $\pi^0$ mesons,
the lateral profile of the CES shower is required to be consistent
with that produced by a single photon. Offline $\pi^{\pm}\gamma$
selection requires each event passing the trigger to have a photon
candidate with $E_T > 25$ \gev\ and a charged pion candidate with $p_T
> 25$ \gevc\ and with significant azimuthal separation ($\Delta \phi > 2$ radians) between
them.

Photon candidates are reconstructed from CEM energy clusters with $|\eta| \le 1.1$ 
and within the CES fiducial. A cluster 
may be pointed to by at most one track; this track must have $p_T$
less than $1\, \gevc + 0.005\times E_T^{\gamma}/c$.  To reject
background from neutral mesons decaying into photons, we veto
candidates which have additional CES clusters with energy above
$2.4\,\gev + 0.01\times E_T^{\gamma}$ in the same CES chamber. As
photons from \wpigamma\ decays are expected to be isolated, we require
the total additional energy deposited in the calorimeter within a cone
of $\Delta R = 0.4$ around the photon candidate to be less than
$1.6\,\gev + 0.02\times E_T^{\gamma}$ and the sum of transverse
momenta of the COT tracks within the same cone to be less than
$2.0\,\gevc +0.005\times E_T^{\gamma}/c$.

Charged pion candidates are reconstructed as COT tracks with impact
parameter $\le 0.2$ cm with respect to the beam axis.  The track must
be consistent with originating from the primary event vertex, which is
required to be within 60 cm of the center of the detector along the
beam line.  The track must point to a narrow calorimeter cluster and
extrapolate to a fiducial region of the CES detector.  As with photons, pions from
\wpigamma\ decays are expected to be isolated, and the pion candidates are
required to satisfy a number of tracking and calorimeter isolation
criteria.  A pion candidate is rejected if there is another track
originating from the same vertex having $p_T > 1$ \gevc\ within a cone
of $10^{\circ}$ around the pion track. The total additional
calorimeter energy within the cone of $\Delta R = 0.4$ around the
cluster is required to be less than 10\% of the track $p_T$ and the sum
of transverse momenta of all tracks within $\Delta R = 0.4$ of the
pion track is required to be less than 5\% of its $p_T$.  To reject
background from QCD jets, which often contain neutral particles, no
CES clusters with energy greater than 500 MeV may be found within
$30^\circ$ of the track.  Finally, pion candidates passing electron
identification criteria are excluded from the analysis. This selection
yields a total of 1398 $\pi^{\pm}\gamma$ candidate events.

The acceptance for the \wpigamma\ decay is determined using a sample
of Monte Carlo (MC) events generated with the \textsc{pythia} event 
generator \cite{pythia} and the CTEQ5L parton distribution 
functions \cite{cteq}.
The angular distribution of pions in the $W$ boson rest frame is described
by the formula $dN/d \cos{\theta} = 1+\cos^{2}\theta$, where $\theta$
is the angle of the pion with respect to the $W$ boson line of flight axis.  This
distribution describes the decay of a spin-one particle ($W$) into a
photon and a pseudoscalar pion averaged over $W$ boson polarizations
and summed over photon polarizations.  The detector response is
simulated with a \textsc{geant3} based simulation program \cite{geant}.

The photon identification efficiency is determined using MC simulated
events and corrected using \zee\ data events, where only probe
electrons with suppressed bremsstrahlung, $|E/p-1|<0.1$, are selected.
The method assumes that the detector response to photons is the same
as that to non-radiating electrons.  Comparing simulations
of electron and photon showers in the CEM, we verified that the
accuracy of this assumption is better than 1\%.

The identification efficiency for pions from the \wpigamma\ decay is
also calculated using MC simulation. The correction factor to this
efficiency, which takes into account the effects of calorimeter shower
mismodeling and instantaneous luminosity, is determined by comparing
properties of the reconstructed jets in photon+jet data to those in
photon+jet MC.

The total corrected acceptance times efficiency for $\pi^{\pm}\gamma$
selection is $A\times\epsilon = 0.0503 \pm 0.0006 (stat) \pm
0.0011 (sys)$.  The uncertainty is dominated by the systematic
uncertainty on the pion identification efficiency.

The dominant backgrounds to this search come from photon+jet
events, where the jet fragments into a single charged particle, and from
multi-jet events, where one of the jets fragments into a single charged
particle and another is misidentified as a photon. Drell-Yan pair
production and $W/Z$ decays, especially to $\tau$ leptons, also contribute
to the background at a level of $\approx 10 \%$.  The signal from the
\wpigamma\ decay would appear as a peak in the $\pi^{\pm}\gamma$ 
invariant mass spectrum centered at the $W$ boson mass with a resolution of
2.5 \gevcc, which includes the experimental resolution and the full
width of the $W$ boson, 2.1 $\gevcc$ \cite{pdg}. We therefore define the
signal region as $75< M_{\pi^{\pm}\gamma} <85$ $\gevcc$, which includes 90\%
of \wpigamma\ decays.  The background within the signal region is
estimated by fitting the sidebands, $67.5 < M_{\pi^{\pm}\gamma} <
75$ \gevcc\ and $85 < M_{\pi^{\pm}\gamma} < 120$ \gevcc, with an exponential
function using a $\chi^2$ minimization.

\begin{figure}[hbtf]
\begin{center}
\includegraphics[width=3.3 in]{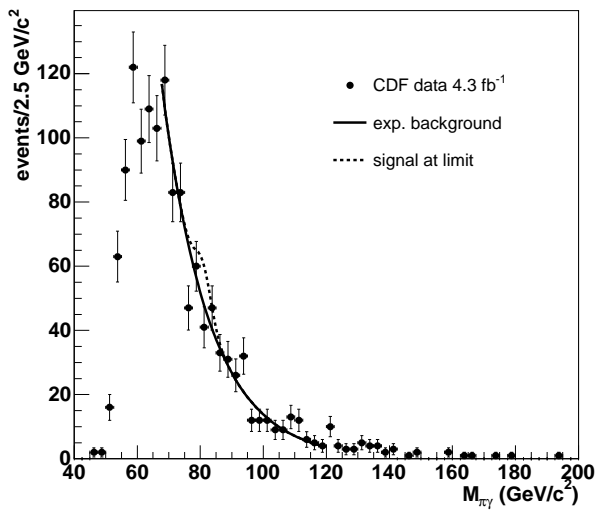}
\caption[] { 
$M_{\pi^{\pm}\gamma}$ and background expectation for 1398 events passing full
selection. The signal expectation at the 95\% C.L. upper limit
is included as the dashed curve. The uncertainties shown are purely statistical.

}
\label{figure:invmass}
\end{center}
\end{figure}

\begin{figure}[hbtf]
\begin{center}
\includegraphics[width=3.3in]{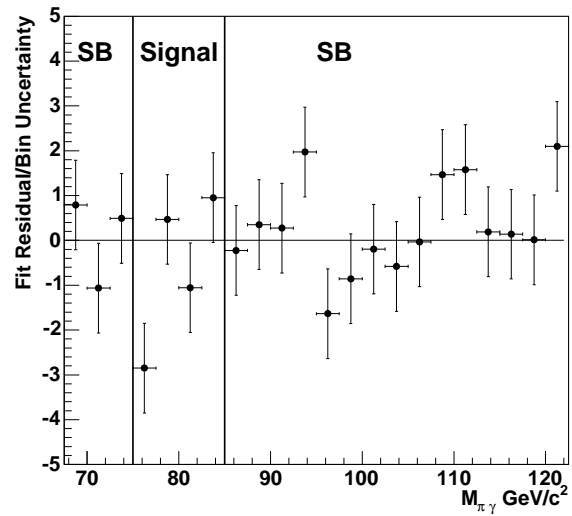}
\caption[] {The fit residual divided by bin uncertainty for Figure\ref{figure:invmass}. The signal and sideband (SB) regions are noted with the vertical lines.	

}
\label{figure:residual}
\end{center}
\end{figure}

Figure \ref{figure:invmass} shows the $M_{\pi^{\pm}\gamma}$ distribution as
well as the sideband fit for all events passing selection. The fit
residuals, plotted in Figure \ref{figure:residual}, show that the
exponential fit is an adequate model of the background shape with the present
level of statistics. The statistical uncertainty on the fit results in a $\approx 5\%$ uncertainty on the background expectation.

From the sideband fit, a total of 219 $\pm$ 10 events are expected
in the signal region from the fit and 206 are observed. Since the data
in the signal region are consistent with the expected
background, we set a 95\% C.L.  upper limit on the
$\sigma(\ppbar\rightarrow W) \times BR(\wpigamma)$.  We divide the
signal region into four 2.5 \gev\ bins.  The limits calculated for
each bin are then combined assuming that uncertainties across the bins
are 100\% correlated. This provides a gain in sensitivity by using
information about the shape of the expected \wpigamma\ mass peak.

For the normalization measurement in the \wenu\ channel we select
events which have a central ($|\eta| < 1.2$) electron with $E_T >
25\,\gev$ and missing transverse energy $\met\ \gt\
25\,\gev$~\cite{MET_defn}.  The electron reconstruction and
identification algorithms used in this analysis are discussed
in \cite{wenu}.  In addition, a standalone measurement of electron
energy in the CES helps to identify electron candidates that radiate a
significant fraction of their energy (due to bremsstrahlung) and
improves overall electron identification efficiency.  This selection
results in a sample with $\le 5\%$ background, the dominant sources of
which are \wtaunu, \zee\ decays and multijet events with one of the
jets misidentified as an electron. The backgrounds are subtracted
using MC simulation and the resulting $\sigma \times BR(\wenu)$ is found to be consistent with previous
measurements \cite{wenu}.  The dominant source of uncertainty on the
cross section is the uncertainty on the luminosity \cite{luminosity},
which cancels in the ratio \bratio.

Other sources of the systematic uncertainties that cancel almost entirely in
the measurement of \bratio\ include the uncertainties on the trigger
efficiency as well as the uncertainties on parton distribution 
functions \cite{pdf} and on the $W$ transverse momentum,
which affect the experimental acceptance. The uncertainties on the
electron and photon identification efficiences cancel partially but
result in a negligible uncertainty on the measurement of \bratio. The
dominant remaining sources of uncertainties are the statistical and
systematic uncertainties on $A\times\epsilon$ ($\sim 2\%$) and the
statistical uncertainty on the $\pi^{\pm}\gamma$ background fit ($\sim
5\%$).

The 95\% C.L. upper limit on the ratio of partial widths
\bratio, calculated using a Bayesian approach that takes into account
the effect of systematic uncertainties \cite{joel}, is $6.4\times
10^{-5}$. This improves the previous world's best limit \cite{kadel}
by more than a factor of 10. The corresponding 95\% C.L. upper limit
on the branching ratio $BR(\wpigamma)$, calculated using the world
average $BR(\wenu)=0.1075\pm 0.0013$\ \cite{pdg}, is $7.0\times
10^{-6}$. While no \wpigamma\ signal was observed, the sensitivity of
the CDF experiment to this decay channel has reached a level comparable to
theoretical predictions. With increases in the dataset size, we expect the 
sensitivity to improve as 1/$\sqrt{L}$, where L is the integrated luminosity.

We thank the Fermilab staff and the technical staffs of the participating institutions for their vital contributions. This work was supported by the U.S. Department of Energy and National Science Foundation; the Italian Istituto Nazionale di Fisica Nucleare; the Ministry of Education, Culture, Sports, Science and Technology of Japan; the Natural Sciences and Engineering Research Council of Canada; the National Science Council of the Republic of China; the Swiss National Science Foundation; the A.P. Sloan Foundation; the Bundesministerium f\"ur Bildung und Forschung, Germany; the Korean World Class University Program, the National Research Foundation of Korea; the Science and Technology Facilities Council and the Royal Society, UK; the Institut National de Physique Nucleaire et Physique des Particules/CNRS; the Russian Foundation for Basic Research; the Ministerio de Ciencia e Innovaci\'{o}n, and Programa Consolider-Ingenio 2010, Spain; the Slovak R\&D Agency; the Academy of Finland; and the Australian Research Council (ARC).


\begin{thebibliography}{99}
\bibitem{MET_defn} 
 Missing \Et\ (\met) is defined by $\vec{\met} = -\Sigma_i
 E_T^i\hat{n}_i$, where $i$ is the calorimeter tower number for $|\eta
 | < 3.6$, and $\hat{n}_i$ is a unit vector perpendicular to the beam
 axis and pointing at the $i^{th}$ tower ($\met = |\vec{\met}|$).
\end{thebibliography}
\end{document}